# New Sensitizing Runs Rules for Shewhart type Control Charts with Applications

Demetrios L. Antzoulakos and Athanasios C. Rakitzis

Department of Statistics and Insurance Science, University of Piraeus, 18534 Piraeus, Greece

**Abstract:** The most popular tool used in the industry for monitoring a process is the Shewhart control chart. The major disadvantage of the Shewhart control chart is that it is not very efficient in detecting small process average shifts. To increase the sensitivity of Shewhart control charts to small shifts additional supplementary runs rules has been suggested. In this paper we introduce and study the modified $r/m$ control chart which has an improved sensitivity to small and moderate process average shifts as compared with the standard Shewhart $\overline{X}$ control chart and corresponding control charts proposed recently in the literature.

## I. INTRODUCTION

The most efficient procedure in statistical process control for monitoring a manufacturing process is the control chart. The Shewhart $\overline{X}$ control chart has been the most popular control chart for monitoring the mean of the distribution (usually a normal distribution) of a quality characteristic of the items produced by a certain process. The standard Shewhart control chart utilizes "three-sigma" control limits (corresponding to 0.27% of false alarms) and indicates an out-of-control signal (shift of the process average) if a single point falls beyond the control limits. Alternative to Shewhart control charts are the CUSUM and EWMA control charts (see, e.g. Montgomery[6]). However, these control charts have not been widely adopted due to their difficulty of use in comparison with the Shewhart control charts which are very easy to design and to interpret.

It is known that Shewhart control charts are efficient in detecting quickly medium to large shifts of the process average, but are insensitive to small shifts. To increase the sensitivity of Shewhart control charts to small shifts additional supplementary runs rules has been suggested and studied (Page[8], Western Electric Company[12], Roberts[10], Bissell[1], Wheeler[13], Nelson[7], Champ and Woodall[2], Palm[9], Shmueli and Cohen[11] and references there in). Traditionally, the performance of a control chart is evaluated by the average run length (ARL). For a specified control chart and a given process average shift ARL is the average number of points plotted on the chart until an out-of-control signal is obtained. Also the waiting time distribution until the occurrence of the signal is called run length distribution. The ARL value associated with a zero (resp. non-zero) process average shift is called in-control (resp. out-of-control) ARL. The standard Shewhart $\overline{X}$ control chart has an in-control ARL equal to 370.4 ($ARL_{in} = 1/0.0027$). A disadvantage of the use of supplementary runs rules in a standard Shewhart $\overline{X}$ control chart is the reduction of the in-control ARL (excessive number o false alarms). The use of the well known Western Electric rules in a Shewhart $\overline{X}$ control chart results in an in-control ARL of 94.75 (see Palm[9]) which is significantly lower from the corresponding value of 370.4 of the standard Shewhart $\overline{X}$ control chart without supplementary runs rules.

In order to compare the ARL performance of various supplementary runs rules they ought to yield the same in-control ARL value. Since this feature is absent from the Shewhart control charts with supplementary runs rules (see, e.g. Champ and Woodall[2]) we are not able to proceed to such a comparison. Recently, Klein[4] suggested two alternatives to the Shewhart $\overline{X}$ control chart: the two of two ($2/2$) scheme and the two of three ($2/3$) scheme. Both control charts are based on runs rules and they have symmetric upper and lower control limits (UCL, LCL). The specific values of the UCL and LCL depend on the in-control ARL value we are willing to work with. Both control charts are easily implemented and have better ARL performance than the standard Shewhart $\overline{X}$ control chart (1/1 scheme) for process average shifts as large as 2.6 standard deviations from the mean. In the same direction, Khoo[5] proceeded to a simulation study of the ARL performance of the $2/2$, $2/3$, $2/4$, $3/3$ and $3/4$ control charts and concluded that the $3/4$ control chart is the most sensitive scheme for detecting small process average shifts.

In the present paper we propose a modified version of the $r/m$ scheme ($r < m$) of Klein[4]. The new resulting control chart, which we call modified $r/m$ control chart, has better ARL performance than the corresponding $r/m$ control charts studied by Klein[4] and Khoo[5]. Also, we have proceeded a detailed study of the new control chart and we give exact ARL and standard deviation values of the modified $r/m$ control chart with the in-control ARL value equal to 370.4 for $m = 1(1)5$ and $r = 1(1)5$ ($r \leq m$). For comparison reasons we give exact ARL and standard deviation values for the $r/m$ control charts of Klein[4] and Khoo[5]. Our study reveals that for process average shifts from 0 to 2.6 standard deviations a modified $r/5$ scheme ($r = 2,3,4$) has always the best overall ARL performance. For the modified $r/5$ scheme ($r = 2,3,4$) we give percentile points of the run length distribution and we compare its performance with the performance of a Shewhart $\overline{X}$ control chart which utilizes the Western Electric rules.

## II. THE MODIFIED $r/m$ CONTROL CHART

Klein[4], motivated by Derman and Ross[3], considered the following two runs rules schemes alternative to the Shewhart $\overline{X}$ control chart: (a) the two of two ($2/2$) scheme which gives an out-of-control signal if either two successive points are plotted above an upper control limit (UCL) or two successive points are plotted below a lower control limits (LCL), and (b) the two of three ($2/3$) scheme which gives an out-of-control signal if two out of three successive points are plotted above an upper control limit or two out of three successive points are plotted below a lower control limit. Klein[4] concluded that both schemes have better ARL performance than the standard Shewhart scheme for detecting process average shifts up to 2.6 standard deviations (the $2/3$ scheme has better performance that the $2/2$ scheme).

Both schemes can be considered as special cases of the $r$ of $m$ ($r/m$) scheme, $1 \leq r \leq m$, which gives an out-of-control signal if either $r$ out of $m$ successive points are plotted above an upper control limit or $r$ out of $m$ successive points are plotted below a lower control limit. The aforementioned schemes (a) and (b) correspond to the $2/2$ and $2/3$ schemes, respectively, while the standard Shewhart control chart corresponds to the $1/1$ scheme. When an out-of-control signal is due to an $r/m$ scheme with $m > r$, we observe $r$ points which lie all above (resp. below) the UCL (resp. LCL), and at most $(m - r)$ points which are placed between the aforementioned $r$ points.

Khoo[5] proceeded to a detailed study of the $2/2$, $2/3$, $2/4$, $3/3$ and $3/4$ schemes and concluded that the $3/4$ scheme is the most sensitive scheme for detecting small process average shifts.

Therefore, we introduce the modified $r$ of $m$ ($M - r/m$) control chart which gives an out-of-control signal if either $r$ points are plotted above an upper control limit that are separated by at most $(m - r)$ points which are placed between the center line and the upper control limit or $r$ points are plotted below a lower control limit that are separated by at most $(m - r)$ points which are placed between the center line and the lower control limit. We note that the $r/r$ and $M - r/r$ schemes are the same schemes. In the following section we proceed to a detailed study of the performance of the $M - r/m$ control chart

## III. PERFORMANCE OF THE MODIFIED $r/m$ CONTROL CHART

Without loss of generality we assume that the random variables giving rise to the points plotted on the control chart are independent and normally distributed with a constant standard deviation equal to one ($\sigma = 1$). The process is considered to be in-control when the process mean is zero ($\mu = 0$).

Table 1 contains ARL values of the $r/r$ schemes for $r = 1(1)5$ and also the $r/m$ schemes and the $M - r/m$ for $r = 2(1)4$, $m = 3(1)5$ and their respective control limits. Process average shifts vary from zero (in-control) to out-of-control values up to four-sigma. The in-control ARL ($ARL_{in}$) for all schemes is set to 370.4 corresponding to the in-control ARL of the standard Shewhart three-sigma $\overline{X}$ control chart (column 1/1 of Table 1). The bold faced ARL values give the lowest ARL value for every process average shift. Since the ARL, as a single parameter, does not contains all of the information on the run length distribution, the standard deviation (SD) of the run length distribution is also given in parentheses.

**Table 1.** ARL and SD values for $r/r$, $r/m$ and $M-r/m$ schemes: $ARL_{in} = 370.40$

| Shift | 1/1 | 2/2 | 3/3 | 4/4 | 5/5 | M-2/3 | 2/3 | M-2/4 | 2/4 | M-3/4 | 3/4 | M-2/5 | M-3/5 | M-4/5 |
|---|---|---|---|---|---|---|---|---|---|---|---|---|---|---|
| | | | | | | Control Limits | | | | | | | | |
| | ± 3 | ± 1.781 | ± 1.2 | ± 0.832 | ± 0.568 | ± 1.866 | ± 1.929 | ± 1.897 | ± 2.011 | ± 1.312 | ± 1.393 | ± 1.91 | ± 1.358 | ± 0.949 |
| 0.0 | 370.40 | 370.40 | 370.40 | 370.40 | 370.40 | 370.40 | 370.40 | 370.40 | 370.40 | 370.40 | 370.40 | 370.40 | 370.40 | 370.40 |
| | (369.90) | (368.94) | (368.03) | (367.13) | (366.27) | (368.63) | (368.47) | (368.04) | (368.43) | (367.61) | (367.44) | (368.28) | (367.30) | (366.68) |
| 0.2 | 308.43 | 276.67 | 259.30 | 248.54 | **241.32** | 264.79 | 270.10 | 257.81 | 266.96 | 243.10 | 248.65 | 253.39 | 233.55 | **231.24** |
| | (307.93) | (275.22) | (256.96) | (245.34) | (237.28) | (263.03) | (268.20) | (264.64) | (255.82) | (240.35) | (245.76) | (251.24) | (230.48) | (227.61) |
| 0.4 | 200.10 | 150.25 | 129.55 | 118.70 | **112.26** | 134.92 | 141.61 | 126.61 | 137.81 | 112.01 | 117.78 | 121.52 | 102.82 | **101.68** |
| | (199.58) | (148.82) | (127.26) | (115.96) | (108.37) | (133.18) | (139.78) | (135.58) | (124.63) | (109.34) | (115.01) | (119.35) | (99.83) | (98.18) |
| 0.6 | 119.67 | 78.91 | 65.25 | 58.99 | **55.71** | 67.89 | 72.64 | 62.24 | 70.12 | 53.79 | 57.48 | 58.85 | **48.26** | 48.34 |
| | (119.16) | (77.51) | (63.02) | (55.98) | (51.95) | (66.18) | (70.86) | (67.99) | (60.29) | (51.21) | (54.83) | (56.70) | (45.37) | (44.98) |
| 0.8 | 71.55 | 43.63 | 35.76 | 32.63 | **31.28** | 36.64 | 39.64 | 33.22 | 38.18 | 28.83 | 31.04 | 31.21 | **25.71** | 26.28 |
| | (71.05) | (42.25) | (33.59) | (29.71) | (27.63) | (34.97) | (37.92) | (36.15) | (31.33) | (26.34) | (28.49) | (29.12) | (22.93) | (23.03) |
| 1.0 | 43.90 | 25.78 | 21.45 | 20.06 | **19.72** | 21.44 | 23.30 | 19.42 | 22.50 | 17.23 | 18.57 | 18.26 | **15.46** | 16.18 |
| | (43.39) | (24.42) | (19.34) | (17.20) | (16.13) | (18.82) | (21.64) | (20.57) | (17.59) | (14.82) | (16.11) | (16.25) | (12.78) | (13.03) |
| 1.2 | 27.82 | 16.28 | 14.00 | **13.54** | 13.72 | 13.56 | 14.73 | 12.37 | 14.30 | 11.36 | 12.18 | 11.70 | **10.32** | 11.09 |
| | (27.32) | (19.94) | (11.92) | (10.73) | (10.18) | (11.99) | (13.12) | (12.45) | (10.60) | (9.00) | (9.80) | (9.77) | (7.72) | (7.98) |
| 1.4 | 18.25 | 10.94 | **9.85** | 9.91 | 10.37 | 9.21 | 9.96 | 8.49 | 9.74 | 8.14 | 8.67 | 8.11 | **7.53** | 8.30 |
| | (17.74) | (9.62) | (7.79) | (7.11) | (6.82) | (7.67) | (8.40) | (7.97) | (6.78) | (5.82) | (6.33) | (6.25) | (4.98) | (5.20) |
| 1.6 | 12.38 | 7.79 | **7.41** | 7.77 | 8.39 | 6.67 | 7.16 | 6.23 | 7.06 | 6.26 | 6.62 | 6.02 | **5.90** | 6.67 |
| | (11.87) | (6.48) | (5.35) | (4.95) | (4.80) | (5.15) | (5.63) | (5.34) | (4.56) | (3.95) | (4.28) | (4.21) | (3.37) | (3.55) |
| 1.8 | 8.70 | **5.85** | 5.89 | 6.44 | 7.16 | 5.10 | 5.43 | 4.84 | 5.40 | 5.11 | 5.35 | **4.72** | 4.91 | 5.69 |
| | (8.18) | (4.54) | (3.82) | (3.58) | (3.49) | (3.60) | (3.92) | (3.72) | (3.19) | (2.78) | (3.01) | (2.96) | (2.36) | (2.50) |
| 2.0 | 6.30 | **4.61** | 4.92 | 5.59 | 6.38 | 4.10 | 4.33 | 3.95 | 4.33 | 4.38 | 4.55 | **3.89** | 4.27 | 5.07 |
| | (5.78) | (3.29) | (2.81) | (2.66) | (2.60) | (2.60) | (2.82) | (2.67) | (2.31) | (2.01) | (2.17) | (2.15) | (1.70) | (1.80) |
| 2.2 | 4.70 | **3.79** | 4.28 | 5.03 | 5.87 | 3.44 | 3.60 | 3.35 | 3.62 | 3.91 | 4.02 | **3.33** | 3.85 | 4.67 |
| | (4.19) | (2.45) | (2.12) | (2.01) | (1.97) | (1.93) | (2.09) | (1.97) | (1.71) | (1.47) | (1.59) | (1.61) | (1.26) | (1.31) |
| 2.4 | 3.65 | **3.23** | 3.85 | 4.66 | 5.54 | 2.99 | 3.10 | 2.95 | 3.14 | 3.59 | 3.68 | **2.94** | 3.57 | 4.42 |
| | (3.11) | (1.87) | (1.63) | (1.54) | (1.50) | (1.45) | (1.57) | (1.49) | (1.30) | (1.10) | (1.18) | (1.24) | (0.95) | (0.96) |
| 2.6 | 2.90 | **2.85** | 3.56 | 4.42 | 5.33 | 2.68 | 2.76 | 2.66 | 2.80 | 3.39 | 3.44 | **2.66** | 3.38 | 4.26 |
| | (2.35) | (1.45) | (1.26) | (1.19) | (1.15) | (1.11) | (1.20) | (1.14) | (1.00) | (0.82) | (0.88) | (0.97) | (0.72) | (0.70) |
| 2.8 | **2.38** | 2.58 | 3.36 | 4.26 | 5.20 | 2.47 | 2.52 | 2.46 | 2.56 | 3.25 | 3.29 | 2.46 | 3.25 | 4.16 |
| | (1.81) | (1.13) | (0.98) | (0.91) | (0.87) | (0.86) | (0.93) | (0.89) | (0.79) | (0.62) | (0.66) | (0.77) | (0.56) | (0.52) |
| 3.0 | **2.00** | 2.39 | 3.23 | 4.16 | 5.11 | 2.32 | 2.36 | 2.32 | 2.39 | 3.16 | 3.18 | 2.32 | 3.16 | 4.09 |
| | (1.41) | (0.89) | (0.76) | (0.70) | (0.66) | (0.67) | (0.72) | (0.70) | (0.62) | (0.46) | (0.50) | (0.62) | (0.44) | (0.38) |
| 3.5 | **1.45** | 2.14 | 3.07 | 4.04 | 5.03 | 2.11 | 2.13 | 2.12 | 2.15 | 3.05 | 3.05 | 2.12 | 3.05 | 4.02 |
| | (0.80) | (0.24) | (0.40) | (0.34) | (0.31) | (0.36) | (0.39) | (0.35) | (0.40) | (0.23) | (0.25) | (0.36) | (0.23) | (0.17) |
| 4.0 | **1.19** | 2.04 | 3.02 | 4.01 | 5.00 | 2.03 | 2.04 | 2.04 | 2.05 | 3.01 | 3.01 | 2.04 | 3.01 | 4.00 |
| | (0.47) | (0.07) | (0.19) | (0.15) | (0.13) | (0.19) | (0.20) | (0.19) | (0.22) | (0.11) | (0.12) | (0.20) | (0.11) | (0.07) |

As expected, the ARL performance of all schemes is better than the standard Shewhart control chart for small to moderate process average shifts. A direct inspection in Table 1, reveals that a $M-r/m$ scheme has better ARL performance that the corresponding $r/m$ scheme for every process average shifts. It also follows that for process average shifts from 0 to 2.6 standard deviations a $M-r/5$ schemes ($r=2,3,4$) has always the best overall ARL performance. However, for larger process average shifts the standard Shewhart control chart performs slightly better than the other schemes. Furthermore, the best ARL performance between an $r/r$ scheme and a $M-k/r$ scheme for $k<r$ (both schemes monitor the last $r$ successive points plotted on the chart) depends on the process average shift and the values of $r$ and $ARL_{in}$.

Since the run length distribution is a highly skewed distribution with a right tail which decreases slowly for small process average shifts, practitioners are more interested in percentiles of the run length distribution (see, Palm[9]) than in the ARL. Therefore, in Tables 2 – 4 we give percentiles points of the run length distribution corresponding to the best ARL performing $M-r/5$ schemes ($r=2,3,4$) for in-control ARL value equal to 370.4.

**Table 2.** Percentiles and ARL values for the $M-2/5$ scheme: $ARL_{in}=370.4$

| Shift | ARL | \multicolumn{5}{c}{Percentiles} |
| | | 5th | 25th | 50th | 75th | 95th |
|---|---|---|---|---|---|---|
| 0.0 | 370.40 | 21 | 108 | 257 | 513 | 1105 |
| 0.2 | 253.30 | 15 | 74 | 176 | 350 | 755 |
| 0.4 | 121.52 | 8 | 37 | 85 | 168 | 360 |
| 0.6 | 58.85 | 5 | 18 | 41 | 81 | 172 |
| 0.8 | 31.21 | 4 | 10 | 22 | 42 | 89 |
| 1.0 | 18.26 | 3 | 7 | 13 | 25 | 51 |
| 1.2 | 11.70 | 2 | 5 | 9 | 15 | 31 |
| 1.4 | 8.11 | 2 | 4 | 6 | 11 | 21 |
| 1.6 | 6.02 | 2 | 3 | 5 | 8 | 14 |
| 1.8 | 4.72 | 2 | 3 | 4 | 6 | 11 |
| 2.0 | 3.89 | 2 | 2 | 3 | 5 | 8 |
| 2.2 | 3.33 | 2 | 2 | 3 | 4 | 6 |
| 2.4 | 2.94 | 2 | 2 | 3 | 3 | 5 |
| 2.6 | 2.66 | 2 | 2 | 2 | 3 | 5 |
| 2.8 | 2.46 | 2 | 2 | 2 | 3 | 4 |
| 3.0 | 2.32 | 2 | 2 | 2 | 3 | 4 |
| 3.5 | 2.12 | 2 | 2 | 2 | 2 | 3 |
| 4.0 | 2.04 | 2 | 2 | 2 | 2 | 2 |

**Table 3.** Percentiles and ARL values for the $M-3/5$ scheme: $ARL_{in}=370.4$

| Shift | ARL | \multicolumn{5}{c}{Percentiles} |
| | | 5th | 25th | 50th | 75th | 95th |
|---|---|---|---|---|---|---|
| 0.0 | 370.40 | 22 | 109 | 258 | 512 | 1103 |
| 0.2 | 233.55 | 15 | 69 | 163 | 323 | 694 |
| 0.4 | 102.82 | 8 | 32 | 72 | 141 | 302 |
| 0.6 | 48.26 | 4 | 16 | 34 | 66 | 139 |
| 0.8 | 25.71 | 4 | 9 | 19 | 35 | 71 |
| 1.0 | 15.46 | 3 | 6 | 11 | 20 | 41 |
| 1.2 | 10.32 | 3 | 5 | 8 | 13 | 26 |
| 1.4 | 7.53 | 3 | 4 | 6 | 9 | 18 |
| 1.6 | 5.90 | 3 | 4 | 5 | 7 | 13 |
| 1.8 | 4.91 | 3 | 3 | 4 | 5 | 10 |
| 2.0 | 4.27 | 3 | 3 | 4 | 5 | 8 |
| 2.2 | 3.85 | 3 | 3 | 3 | 4 | 6 |
| 2.4 | 3.57 | 3 | 3 | 3 | 4 | 6 |
| 2.6 | 3.38 | 3 | 3 | 3 | 4 | 5 |
| 2.8 | 3.25 | 3 | 3 | 3 | 3 | 4 |
| 3.0 | 3.16 | 3 | 3 | 3 | 3 | 4 |
| 3.5 | 3.05 | 3 | 3 | 3 | 3 | 3 |
| 4.0 | 3.01 | 3 | 3 | 3 | 3 | 3 |

**Table 4.** Percentiles and ARL values for the $M-4/5$ scheme: $ARL_{in}=370.4$

| Shift | ARL | \multicolumn{5}{c}{Percentiles} |
| | | 5th | 25th | 50th | 75th | 95th |
|---|---|---|---|---|---|---|
| 0.0 | 370.40 | 23 | 109 | 258 | 512 | 1102 |
| 0.2 | 231.24 | 15 | 69 | 161 | 319 | 685 |
| 0.4 | 101.68 | 9 | 32 | 72 | 140 | 298 |
| 0.6 | 48.34 | 6 | 16 | 35 | 66 | 138 |
| 0.8 | 26.28 | 5 | 10 | 19 | 35 | 72 |
| 1.0 | 16.18 | 4 | 7 | 12 | 21 | 42 |
| 1.2 | 11.09 | 4 | 5 | 9 | 14 | 27 |
| 1.4 | 8.30 | 4 | 5 | 6 | 10 | 19 |
| 1.6 | 6.67 | 4 | 4 | 5 | 8 | 14 |
| 1.8 | 5.69 | 4 | 4 | 5 | 6 | 11 |
| 2.0 | 5.07 | 4 | 4 | 4 | 5 | 9 |
| 2.2 | 4.67 | 4 | 4 | 4 | 5 | 8 |
| 2.4 | 4.42 | 4 | 4 | 4 | 5 | 6 |
| 2.6 | 4.26 | 4 | 4 | 4 | 4 | 5 |
| 2.8 | 4.16 | 4 | 4 | 4 | 4 | 5 |
| 3.0 | 4.09 | 4 | 4 | 4 | 4 | 5 |
| 3.5 | 4.02 | 4 | 4 | 4 | 4 | 4 |
| 4.0 | 4.00 | 4 | 4 | 4 | 4 | 4 |

To increase the sensitivity of a Shewhart control chart to detect small changes in the mean it is common in practice to use the four rules (Rules 1-4) introduced by the Western Electric Company (see, e.g. Montgomery[6]). The use of a Shewhart control chart along with the Western Electric rules (to be denoted by $C_{1234}$) results in an in-control ARL of 94.75 (see, Palm[9]). The ARL performance of the $C_{1234}$ and the $M-r/5$ ($r=2,3,4$) control charts is given in Table 5. The semi-interquantile range (SIR), as a measure of the spread of the run length distribution is also given in parentheses. The values corresponding to the $C_{1234}$ scheme were taken form Palm[9].

**Table 5.** ARL and SIR values for $C_{1234}$ and $M-r/5$ schemes: $ARL_{in} = 94.75$

|  | $C_{1234}$ | $M$-2/5 | $M$-3/5 | $M$-4/5 |
|---|---|---|---|---|
|  | Control Limits | | | |
| Shift | ±3 | ±1.57098 | ±1.04853 | ±0.652948 |
| 0.0 | 94.57 | 94.57 | 94.57 | 94.57 |
|  | (50.00) | (50.50) | (50.50) | (50.00) |
| 0.2 | **66.99** | 72.28 | 69.51 | 69.96 |
|  | (34.50) | (38.50) | (36.50) | (36.50) |
| 0.4 | **36.54** | 41.51 | 38.27 | 39.16 |
|  | (18.00) | (22.00) | (19.50) | (19.50) |
| 0.6 | **20.88** | 23.62 | 21.65 | 22.65 |
|  | (10.00) | (12.00) | (10.50) | (10.50) |
| 0.8 | **13.24** | 14.45 | 13.51 | 14.49 |
|  | (5.50) | (7.00) | (6.00) | (6.50) |
| 1.0 | **9.22** | 9.59 | 9.28 | 10.22 |
|  | (3.50) | (4.50) | (3.50) | (4.00) |
| 1.2 | 6.89 | **6.86** | 6.92 | 7.82 |
|  | (2.00) | (3.00) | (2.50) | (3.00) |
| 1.4 | 5.42 | **5.22** | 5.53 | 6.40 |
|  | (2.00) | (1.50) | (2.00) | (2.00) |
| 1.6 | 4.41 | **4.20** | 4.66 | 5.52 |
|  | (1.00) | (1.50) | (1.00) | (1.00) |
| 1.8 | 3.68 | **3.53** | 4.10 | 4.96 |
|  | (1.50) | (1.00) | (1.00) | (0.50) |
| 2.0 | 3.13 | **3.07** | 3.74 | 4.61 |
|  | (1.00) | (1.00) | (0.50) | (0.50) |
| 2.2 | **2.70** | 2.75 | 3.49 | 4.38 |
|  | (1.00) | (0.50) | (0.50) | (0.00) |
| 2.4 | **2.35** | 2.53 | 3.32 | 4.23 |
|  | (0.50) | (0.50) | (0.00) | (0.00) |
| 2.6 | **2.07** | 2.36 | 3.21 | 4.14 |
|  | (1.00) | (0.50) | (0.00) | (0.00) |
| 2.8 | **1.85** | 2.25 | 3.13 | 4.08 |
|  | (0.50) | (0.00) | (0.00) | (0.00) |
| 3.0 | **1.67** | 2.17 | 3.08 | 4.05 |
|  | (0.50) | (0.00) | (0.00) | (0.00) |

The results of the above table reveal that the use of the simple $M-r/5$ control chart for detecting small to moderate process average shifts suggested in the present article leads to ARL values very close to the ones achieved by the more complicated $C_{1234}$ control chart.

## ACKNOWLEDGMENT

The work of Athanasios Rakitzis is supported by the National Scholarship Foundation of Greece.